# Atomic Mechanism of Flow in Simple Liquids under Shear

T. Iwashita[1] and T. Egami[1,2,3]

[1]*Joint Institute for Neutron Sciences and Department of Physics and Astronomy, University of Tennessee, Knoxville, TN 37996*

[2]*Department of Materials Science and Engineering, University of Tennessee, Knoxville, TN 37996*

[3]*Oak Ridge National Laboratory, Oak Ridge, TN 37831 USA*

Atomic correlations in a simple liquid in steady-state flow under shear stress were studied by molecular dynamics simulation. The local atomic level strain was determined through the anisotropic pair-density function (PDF). The atomic level strain has a limited spatial extension whose range is dependent on the strain rate and extrapolates to zero at the critical strain rate. A failure event is identified with altering the local topology of atomic connectivity by exchanging bonds among neighboring atoms.

PACS #: 83.50.Ax, 83.10.Rs, 66.20.Cy, 2b,

The flow of a liquid has been theoretically described by the hydrodynamic theories [1], and more recently by the non-linear extension of the hydrodynamic theory, such as the mode-coupling theory [2]. Much less attention has been paid to the atomic level dynamics, because it has been believed that the atomic motion in a liquid was so random that details of the atomic motion were irrelevant to the physics of liquid flow. In the meantime molecular dynamics (MD) simulations of model liquids revealed interesting collectivity of atomic motion, generally known as dynamic heterogeneity, particularly in the liquid state below the melting temperature, the supercooled state [3-7]. Experimental results on diffusion also suggest highly collective mechanisms of atomic transport [8]. However, the subject is still very controversial. In this letter using MD simulations of a simple liquid at relatively low temperatures we show that at the atomic level dynamics of a liquid under steady-state shear flow is highly heterogeneous both in space and time, and the structure of a flowing liquid is strained over a limited spatial range. The spatial extension of the strained region is dependent on the shear rate. The flow is triggered by local alteration of the topology of atomic connectivity. We show that the Maxwell relaxation time, $\eta/G_\infty$, where $\eta$ is viscosity and $G_\infty$ is the instantaneous shear modulus, is directly related to the lifetime of the local topology.

The MD simulations of steady-state flow were performed for a monatomic model consisting of $N = 16,384$ atoms which interacted via a modified Johnson potential for iron which has a minimum at $a = 2.61$ Å [9]. The MD time step was $10^{-15}$ sec. and the number density of atoms was kept constant at $\rho_0 = 0.07843$ Å$^{-3}$. We solved the SLLOD equations of motion [10] with a Gaussian thermostat under the Lees-Edward boundary conditions. Initially a glassy state was prepared by quenching the system from a liquid state at high temperature. Then flow was imposed in the $x$ direction with the velocity gradient in the $z$ direction with various shear strain rates, $\dot{\gamma}$, at temperatures below $T_g$ which is about 900 K in the computer time scale [11]. Even at temperatures below $T_g$ a liquid state is attained if sufficient stress is applied [12,13]. Steady-state flow was achieved after a transient behavior. In the SLLOD algorithm the external force is applied to each particle, avoiding the development of macroscopically inhomogeneous flow, such as formation of shear bands. We chose the temporal unit to be $\tau_0 = \sqrt{ML^2/E}$, where $M$ is the mass of Fe, $L = 1$ Å and $E = 1$ eV, thus $\tau_0 = 0.760845 \times 10^{-13}$ sec. The ranges of strain rates were $2.5\times10^{-5}\ \tau_0^{-1} \leqq \dot{\gamma} \leqq 0.1\ \tau_0^{-1}$ for 300 K, $5.0\times10^{-5}\ \tau_0^{-1} \leqq \dot{\gamma} \leqq 0.1\ \tau_0^{-1}$ for 500 K, $1\times10^{-4}\ \tau_0^{-1} \leqq \dot{\gamma} \leqq 0.2\ \tau_0^{-1}$ for 700 K, and $1\times10^{-3}\ \tau_0^{-1} \leqq \dot{\gamma} \leqq 0.24\ \tau_0^{-1}$ for 900 K.

The atomic structure of the liquid during the steady state shear flow can be described by the same-time (snapshot) atomic pair-density function (PDF),

$$\rho_0 g(\mathbf{r}) = \frac{1}{N\tau} \sum_{i,j} \int_0^\tau \delta\left(\mathbf{r} - \left(\mathbf{r}_i(t) - \mathbf{r}_j(t)\right)\right) dt \quad (1)$$

where $\tau$ is the total simulation time which is much longer than the time scale for atomic dynamics, and $\mathbf{r}_i(t)$ is the position of the *i*-th atom at time *t*. It is generally assumed that the apparent structural anisotropy in a liquid under steady state flow is small, because the PDF does not show significant changes [14, 15]. However, the fact a liquid sustains stress through viscosity implies some structural anisotropy should exist in the liquid [16]. In order to describe such structural anisotropy we use the anisotropic PDF [17-21],

$$\rho_0 g(\mathbf{r}) = \rho_0 \sum_{\ell,m} g_\ell^m(r) Y_\ell^m\left(\frac{\mathbf{r}}{|\mathbf{r}|}\right) \quad (2)$$

where $Y_\ell^m(z)$ are the spherical harmonics and $g_\ell^m(r)$ are the anisotropic PDF. Further details of this approach are described in Supplementary Material. In order to confirm that the observed features are not potential-dependent we also carried out similar MD simulations on a binary model loosely representing $Cu_{50}Zr_{50}$ [22]. The results on the binary model were very similar, as will be reported elsewhere. At high shear rates both the monoatomic and binary systems crystallized forming two-dimensional crystalline sheets. Here we report only the behavior of the monoatomic system before crystallization.

The isotropic PDFs, $g_0^0(r)$, at $T = 300$ K and at various shear strain rates are shown in Fig. 1 (a). Even though temperature is kept at 300 K, $g_0^0(r)$ loses its fine features as the strain rate is increased, and at high strain rates it strongly resembles that of a liquid. This result supports the idea that the stress and temperature play similar roles, and high stress can induce glass transition even at low temperatures [12,13]. The anisotropic PDF relevant for shear deformation in the *z-x* plane is,

$$g_{zx}(r) = \frac{1}{\sqrt{2}}\left[g_2^{-1}(r) - g_2^1(r)\right]. \tag{3}$$

The anisotropic PDF, $g_{zx}(r)$, at $T = 300$ K and $\dot{\gamma} = 0.0001 \times \tau_0^{-1}$ is shown in Fig. 1 (b). Although the magnitude of $g_{zx}(r)$ is smaller than that of $g_0^0(r)$, the presence of significant structural anisotropy under shear is evident. The variations in $g_{zx}(r)$ with strain rate are shown in Supplementary Material.

The system is under the macroscopic stress because of viscosity. At low strain rates the system is very viscous, and over a short period of time it may behave like a glass elastically deformed under shear. Therefore it is conceivable that the instantaneous (same time) structure of a liquid under steady state flow shows some resemblance to that of an elastically deformed glass. For homogeneous (affine) deformation $g_{zx}(r)$ is proportional to the derivative of $g_0^0(r)$ [16,17],

$$g_{zx}^{affine}(r) = -\frac{\gamma_{zx}}{\sqrt{15}} r \frac{d}{dr} g_0^0(r) \tag{4}$$

where $\gamma_{zx}$ is the elastic strain. Indeed the observed $g_{zx}(r)$ closely resembles $g_{zx}^{affine}(r)$ as shown in Fig. 1 (b). However, we found that in order to describe $g_{zx}(r)$ accurately with equation (4) we had to make the strain, $\gamma_{zx}$, dependent on the distance $r$. Equation (4) was fit to the data at each $r$ point over the rage of $r \pm 1.3$ Å to determine the $r$-dependent local strain, $\gamma_{zx}(r)$, as shown in Fig. 2 (a) for various strain rates at $T = 300$ K. The values of $\gamma_{zx}(r)$ are non-zero only over a limited range of distances, and quickly decay to zero beyond this range.

To describe the characteristic length scale of the sheared zone we define the correlation length, $\xi$, measured from the nearest neighbor position at $a = 2.61$ Å, as the distance where $\gamma_{zx}(r)$ decays to half the value of the maximum strain. The correlation length $\xi$ depends on the strain rate $\dot{\gamma}$ as shown in Fig. 2 (b). Interestingly the dependence can be described by a simple function,

$$\dot{\gamma} = \dot{\gamma}_c \exp\left(-(\xi/\xi_T)^2\right), \tag{5}$$

with $\dot{\gamma}_c = 0.30 \times \tau_0^{-1} = 0.394 \times 10^{13}$ sec$^{-1}$. The parameter $\xi_T$ is only weakly dependent on temperature; $\xi_T$ = 10.56 Å at 300 K, 10.85 Å at 500 K, 11.06 Å at 700 K, and 11.12 Å at 900 K. Note that according to Equation (5) $\xi$ extrapolates to zero at the critical strain rate, $\dot{\gamma}_c$. A possible reason for the existence of the critical strain rate is discussed below.

The macroscopic stress, $\sigma$, increases slowly with the strain rate and is weakly temperature dependent as shown in Supplementary Material. But when $\sigma$ is plotted against $\xi$ temperature dependence disappears and it follows a simple relation, $\sigma = \sigma_c \exp(-(\xi/\xi_\sigma))$, with $\sigma_c$ = 0.1439 eV/Å$^3$ and $\xi_\sigma$ = 10.05 Å as shown in Fig. 2 (c). The instantaneous shear strain is given by $\gamma = \sigma/G_\infty$, where $G_\infty$ is the instantaneous shear modulus. In Fig. 3 $\gamma$ is compared with the atomic-level strain at the nearest neighbor, $\gamma_{zx}^{NNS} = \gamma_{zx}(a)$, as a function of $\dot{\gamma}$ at 300 K. They are similar except at very high strain rates, indicating that the atomic-level strain at the nearest neighbor determined by the anisotropic PDF indeed represents the macroscopic instantaneous strain.

We also studied the lifetime of local topology of the atomic structure, or the time it takes for an atom to lose or gain one nearest neighbor, $\tau_{LT}$. Here the nearest neighbor is defined by the minimum in the isotropic PDF, $g_0^0(r)$, between the first and the second peaks, $r_{min}$; if the distance between a pair of atoms is less than $r_{min}$, they are defined as the nearest neighbors to each other. As shown in Supplementary Material we found that the distribution of the time for an atom to lose or gain one nearest neighbor are almost identical, and $\tau_{LT}$ is nearly inversely related to the strain rate. Now, because the system is continuously sheared with the strain rate of $\dot{\gamma}$, the strain built up after the time interval of $\tau_{LT}$ is $\gamma_{LT} = \tau_{LT}\dot{\gamma}$. If we assume for the sake of simplicity that the strain is totally released by losing or gaining one nearest neighbor and accumulates linearly afterwards, resulting in a saw-tooth time dependence of strain, the time average of the strain is equal to $\gamma_{LT}/2$. Surprisingly $\gamma_{LT}/2$ is almost exactly equal to $\gamma_{zx}^{NNS}$ and $\gamma$ as shown in Fig. 3. Because these three quantities, $\gamma_{LT}/2$, $\gamma_{zx}^{NNS}$ and $\gamma$, were determined totally independently, this agreement is important. Now viscosity is given by $\eta = \sigma/\dot{\gamma} = G_\infty \tau_M$, where $\tau_M$ is the Maxwell structural relaxation time. Therefore, the results in Fig. 3 imply,

$$\tau_M = \sigma/G_\infty\dot{\gamma} = \gamma/\dot{\gamma} \approx \gamma_{LT}/2\dot{\gamma} = \tau_{LT}/2, \qquad (6)$$

and conclusively identifies the deformation event with an atom losing or gaining one nearest neighbor. By losing or gaining one nearest neighbor the local topology around an atom is changed, and the system moves from one minimum to another in the energy landscape.

It is interesting to note that events of bond breaking and formation are strongly correlated. When a bond between $i$ and $j$ is broken, a new bond is formed almost immediately among the common neighbors of $i$ and $j$. Fig. 4 shows the distribution of the delay time after the bond breaking and the bond formation among the common neighbors. Its time-scale, about 1.5 $\tau_0$ (= $1.14 \times 10^{-13}$ sec.) is shorter than $\tau_{LT}$ by two orders of magnitude, and is comparable to the time for the shear wave to travel one atomic distance, $a/c = 0.87 \times 10^{-13}$ sec., where $c$ is the shear wave velocity, $3.0 \times 10^3$ m/sec [23]. The inset of Fig. 4 shows a typical change in the nearest neighbor configuration when one bond is broken and a new bond is formed in the immediate neighborhood. Thus bond breaking and formation are elastically coupled as one action of bond-exchange, which causes shear deformation as observed for anelastic deformation by experiment and simulation [18,24]. This event is also the basis for the shear transformation zone (STZ) theory [25] and the asymmetry model [26].

At the critical strain rate $\dot{\gamma}_c$, the time it takes to reach the maximum local strain, $\tau_{LT}^{MAX} = \gamma_{LT}^{MAX}/\dot{\gamma}$, is equal to $0.51 \times 10^{-13}$ sec. This value is of the same order of magnitude as $a/c$, therefore it is most likely that $\dot{\gamma}_c$ is related to a sonic barrier. Because local structural changes cannot propagate faster than the sound, at high strain rates before the effect of the structural change at one atom propagates to the neighbors the atomic cage of the neighbor collapses. Thus local elastic correlations cannot be established at very high strain rates and $\xi$ approaches zero.

In conclusion we studied the atomic correlations in a simple liquid during the steady state flow under shear stress. It is generally assumed that the atomic mechanism of deformation involves an atom leaving the atomic cage. Our study indicates that deformation occurs when the local topology of the cage is changed by an atom losing or gaining just one nearest neighbor, prompting the system to move from one minimum of energy landscape to the other.

**Acknowledgements**
We thank J. S. Langer, A. P. Sokolov, V. Novikov, V. A. Levashov and J. R. Morris for useful discussion, and U. Buchenau for communicating his results prior to publication. The work was supported by the U.S. Department of Energy, Office of Basic Energy Sciences, Materials Science and Engineering Division.

supported by the U.S. Department of Energy, Office of Basic Energy Sciences, Materials Science and Engineering Division.

**Figure captions:**

Figure 1. (a) Isotropic PDF, $g_0^0(r)$, at $T = 300$ K and at various shear strain rates shown in the unit of $\tau_0^{-1} = 1.3143 \times 10^{13}$ sec$^{-1}$. The vertical axis is shifted for clarity. (b) Anisotropic PDF, $g_{zx}(r)$, at $T = 300$ K and $\dot{\gamma} = 0.0001\ \tau_0^{-1}$ (red line) and $g_{zx}^{affine}(r)$ predicted for affine deformation for the elastic strain of $\gamma_{zx} = 0.051$ (dashed black line). $g_{zx}^{affine}(r)$ fits the data well only above 14 Å, and requires an $r$-dependent effective shear strain, $\gamma_{zx}(r)$, to achieve a fit over the entire range.

Figure 2. (a) The variation of the $r$-dependent effective shear strain, $\gamma_{zx}(r)$, with the strain rate, at $T = 300$ K. (b) The correlation length, $\xi$, as a function of the strain rate, $\dot{\gamma}$, for various temperatures. Here $\xi_T = 10.56$ Å at 300 K, 10.85 Å at 500 K, 11.06 Å at 700 K, and 11.12 Å at 900 K. (c) The macroscopic stress, $\sigma$, as a function of $\xi$ at various temperatures. The results are independent of temperature, and follow the exponential form, $\sigma = \sigma_c \exp(-(\xi/\xi_\sigma))$, (solid line).

Figure 3. The atomic-level shear strain at the nearest neighbor, $\gamma_{zx}^{NNS}$, and the macroscopic instantaneous shear strain, $\gamma = \sigma/G_\infty$, are compared with the time-averaged accumulated strain, $\gamma_{LT}/2 = \tau_{LT}\dot{\gamma}/2$. Agreement of the three quantities indicates that the deformation event is identified with cutting of the atomic bond.

Figure 4. Distribution of the time between a bond breaking and bond formation among the common nearest neighbors, $\tau_{delay}$, in the unit of $\tau_0$. The inset shows a typical change in the nearest neighbor configuration when one bond (A-B) is broken and a new bond (C-D) is formed in the immediate neighborhood.

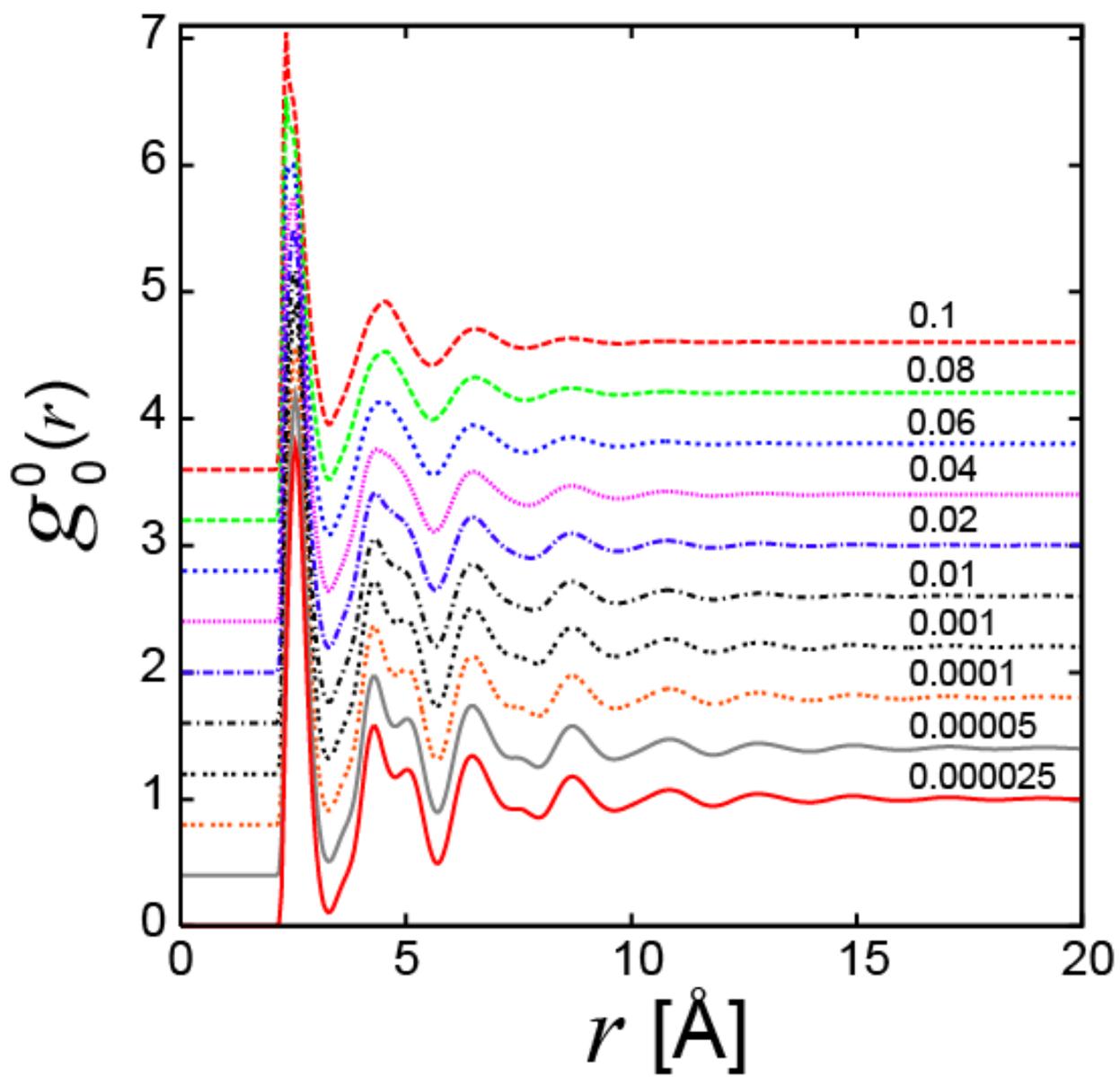

Fig. 1 (a)

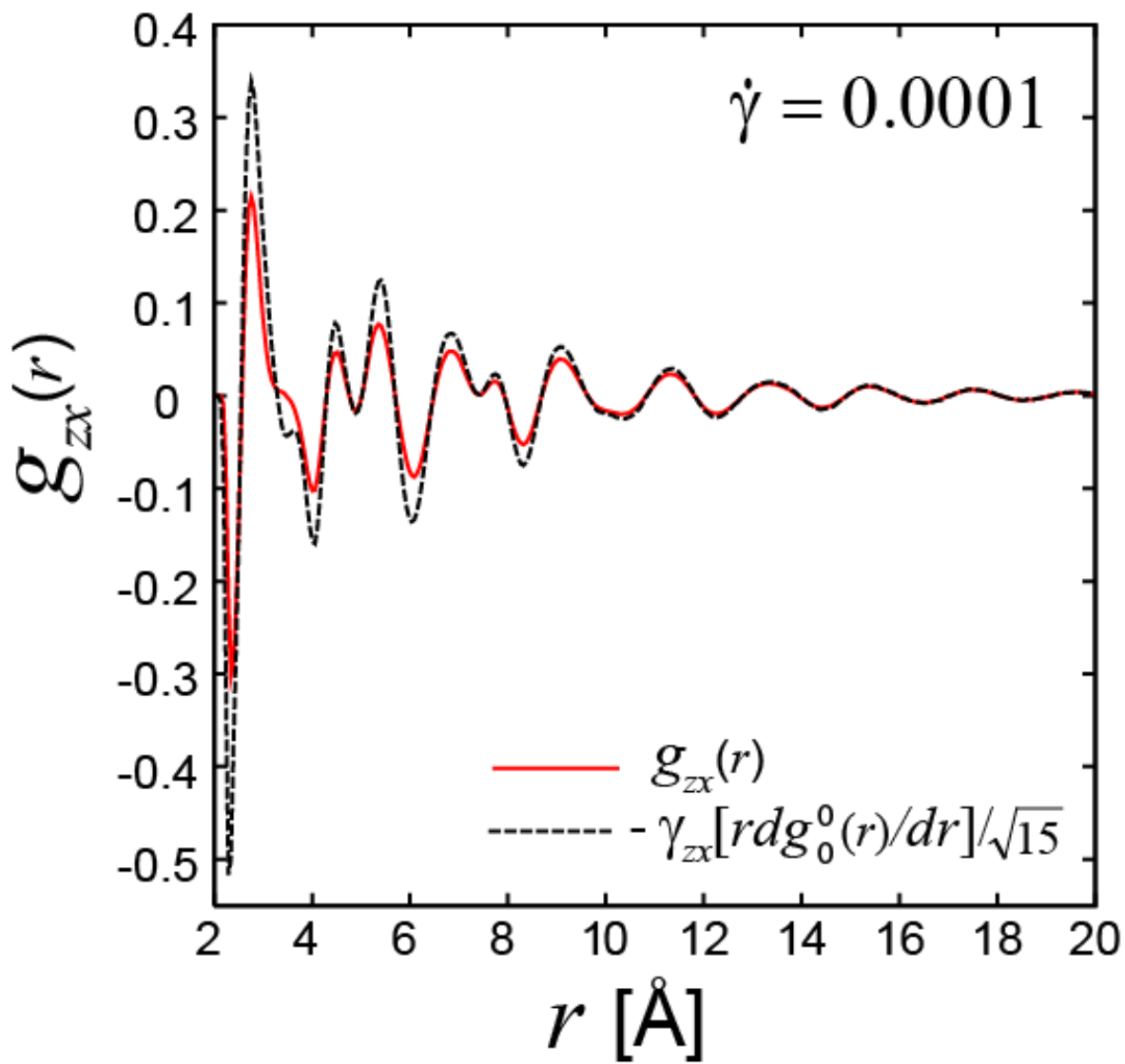

Fig. 1 (b)

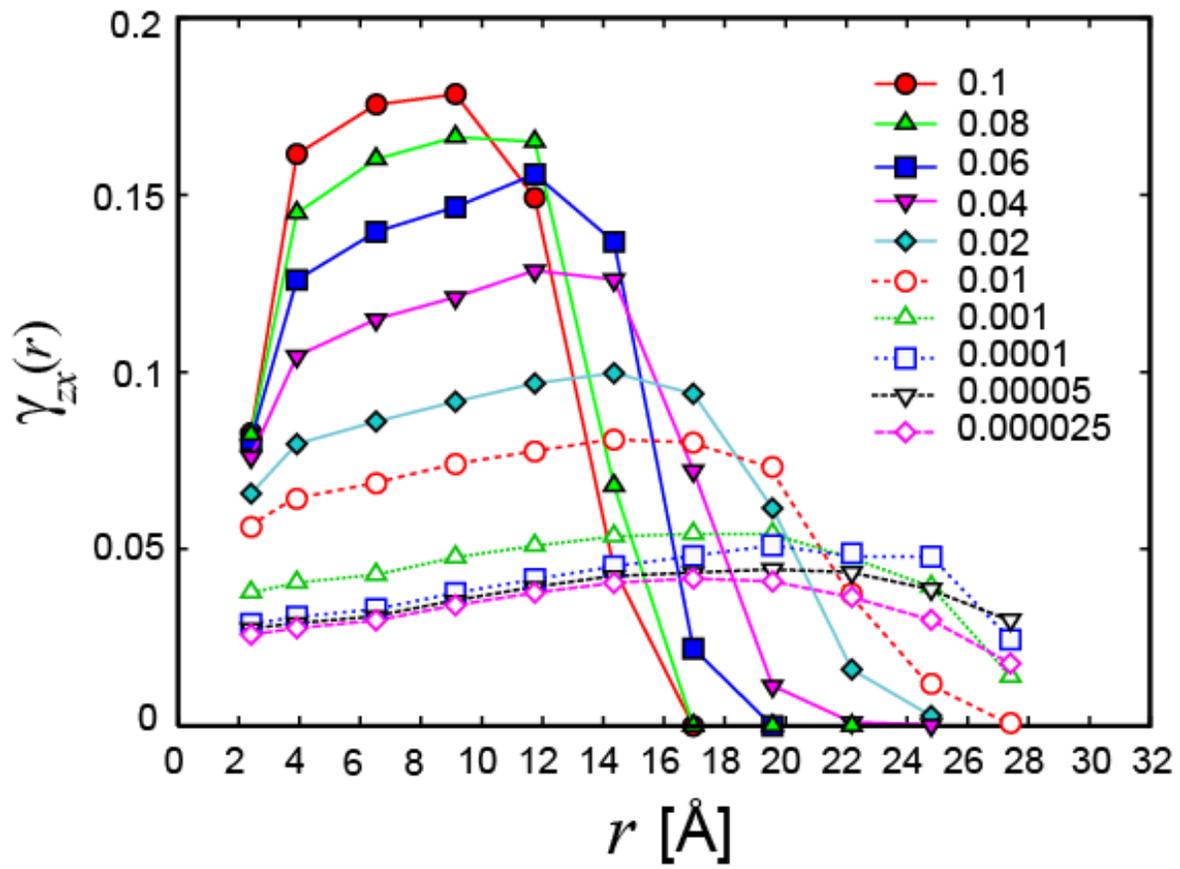

Fig. 2 (a)

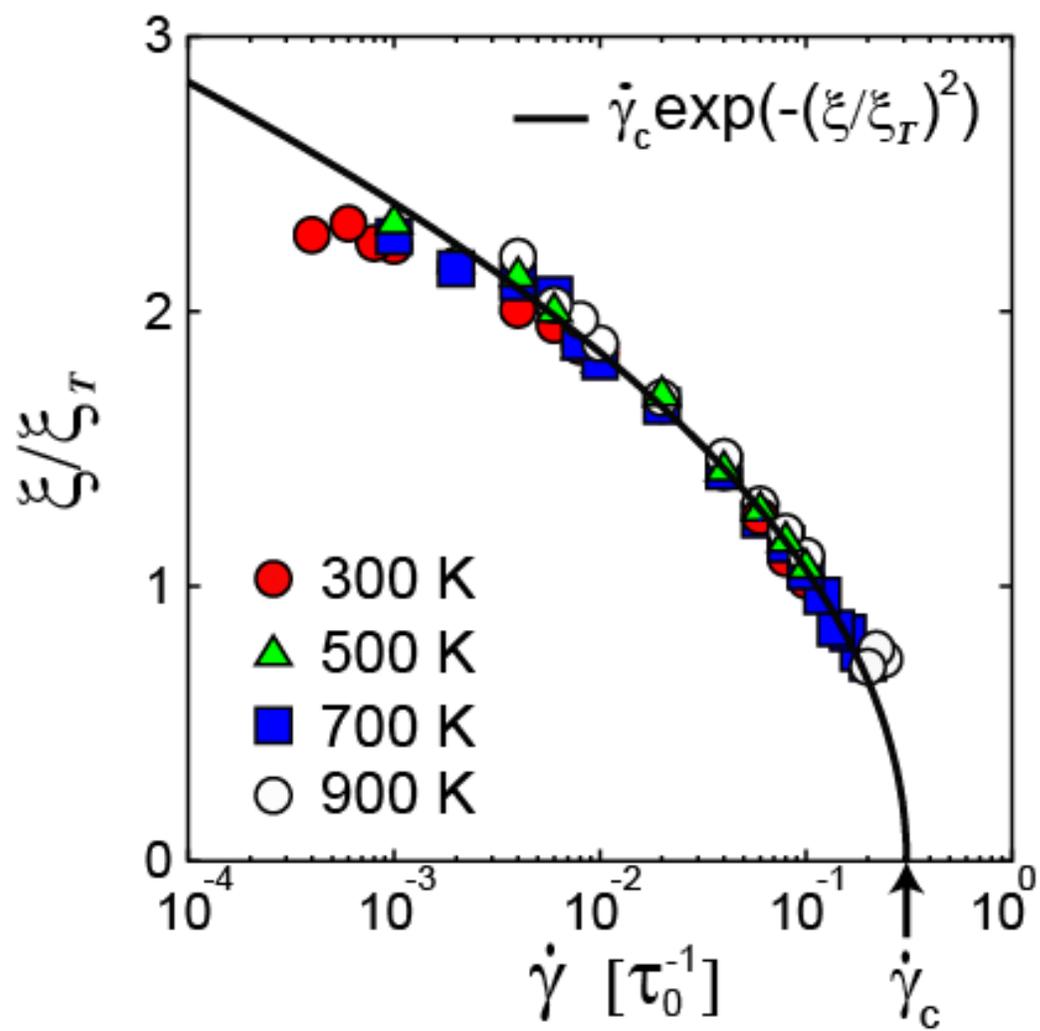

Fig. 2 (b)

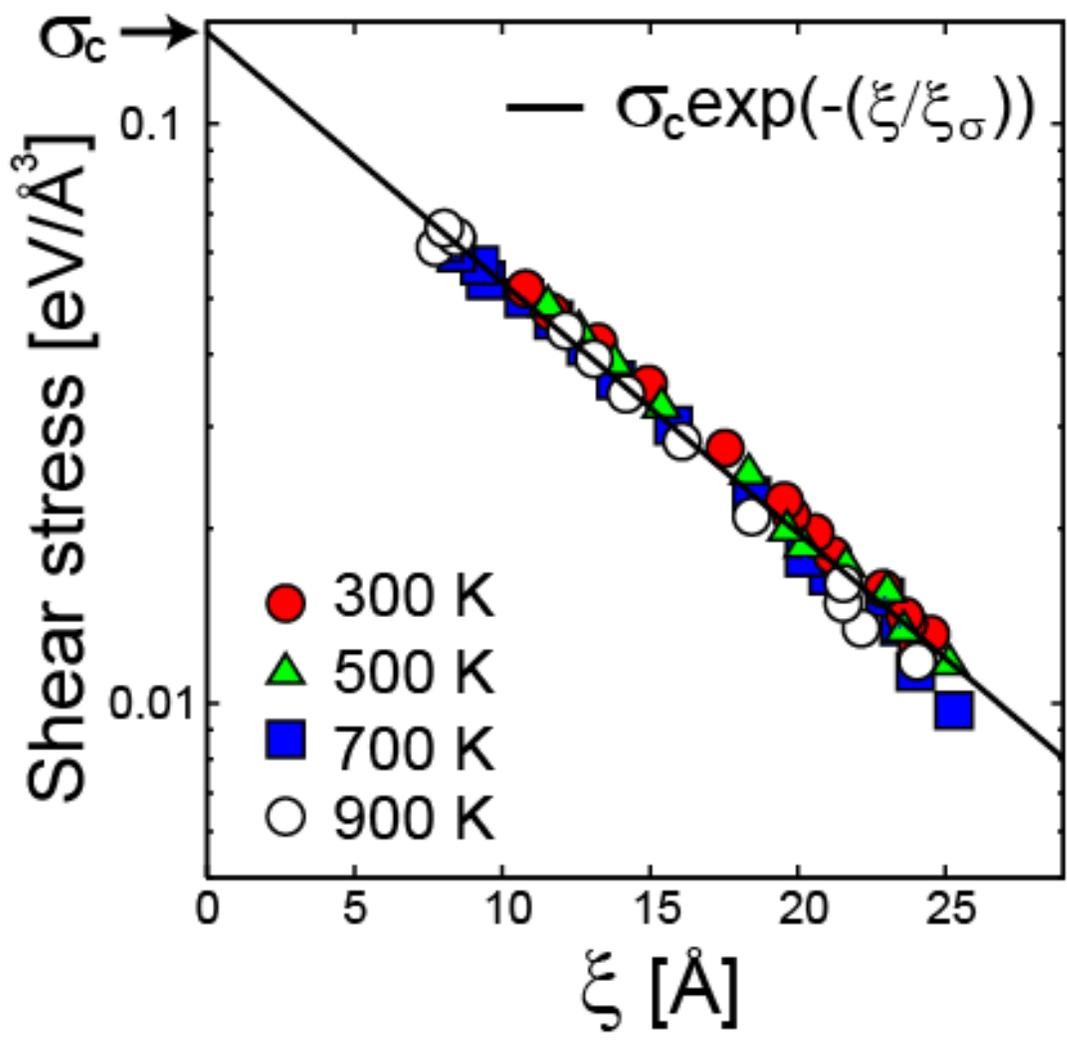

Fig. 2 (c)

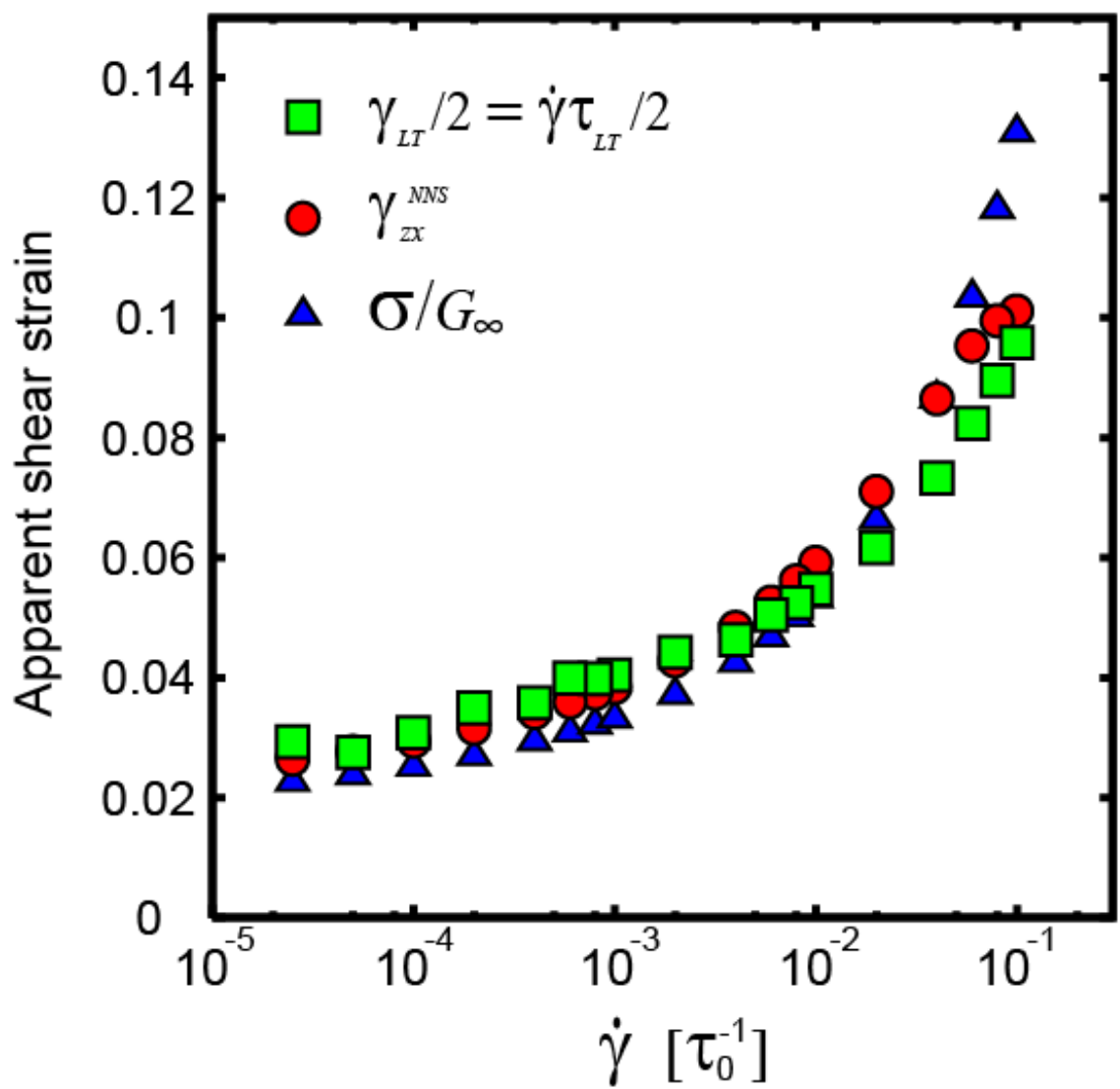

Fig. 3

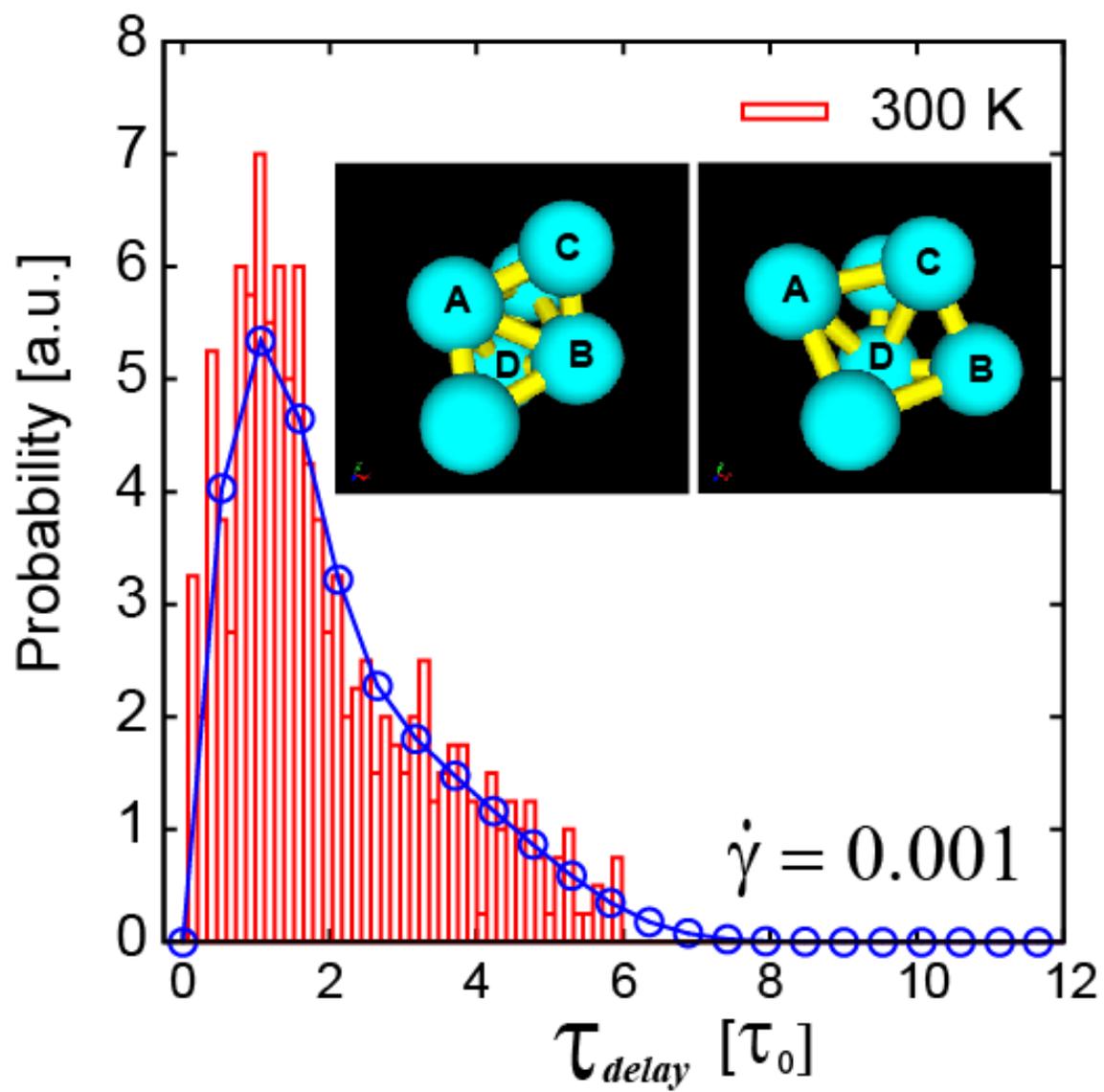